# High- fidelity tracking of the evolution of multilevel quantum states

Yu.I. Bogdanov[*], N.A. Bogdanova, Yu.A. Kuznetsov, V.F. Lukichev

Valiev Institute of Physics and Technology of Russian Academy of Sciences, Moscow, Russia

**ABSTRACT**

The method of quantum tomography, which allows us to track with high accuracy the evolution of multilevel quantum systems (qudits) in Hilbert spaces of various dimensions is presented. The developed algorithms for quantum control are based on the use of the spinor representation of the Lorentz transformation group. In the simplest case of one-qubit states, it turns out that, in addition to three-dimensional rotations on the Bloch sphere, one can introduce four-dimensional Lorentz pseudorotations, similar to the transformations of the special theory of relativity. We show that feedback through weakly perturbing adaptive quantum measurements turns out to be capable of providing high-precision control of the quantum system, while introducing only weak perturbations into the initial quantum state. It turns out that, together with the control of a quantum system through its weak perturbation, the developed algorithms for controlling the evolution of the state of a quantum system can be super-efficient, providing a higher measurement accuracy than any standard POVM (Positive-Operator Valued Measure) protocols. The results of the study are important for the development of optimal adaptive methods for quantum states and operations controlling.

The results obtained are essential for the development of high-precision control methods for quantum information technologies.

## 1. INTRODUCTION

In this paper, we present the quantum tomography method that allows the high-precision evolution of multilevel quantum systems (qudits) in Hilbert spaces of various dimensions. The developed algorithms for quantum control are based on the use of the spinor representation of the group of Lorentz transformers [1,2], as well as its generalizations to multilevel quantum systems [3].

For the case of one-qubit states, we can introduce four-dimensional Lorentz pseudo-rotations in addition to three-dimensional rotations on the Bloch sphere. Such transformations are characteristic of the special theory of relativity. Below we describe an algorithm that provides highly accurate feedback control of a quantum system based on weakly disturbing adaptive quantum measurements. We will show that algorithms for controlling the evolution of the state of a quantum system can be super-efficient, providing higher measurement accuracy than any standard POVM (Positive-Operator Valued Measure) protocols.

In Section 2, spinor and tensor representations of the Lorentz group are considered. In this case, one-qubit density matrices correspond to the spinor representation, and the Minkowski-Stokes vector correspond to the tensor representation. The same section provides basic information about quantum measurement protocols and the accuracy they can provide using quantum tomography procedures. In Section 3, the adaptive tracking of the evolution of a quantum state is considered using the example of an eight-level qudit. Here, in the case of the Lorentz protocols, only a small part of the total number of ions turns out to be strongly disturbed, emitting photons and carrying information about the current state of the qudits. The rest of the ions do not emit photons, but at the same time they are still subject to some weak perturbation (such qudits provide a resource for step-by-step tracking of the evolution of a quantum state). Section 4 presents the main findings of the study.

## 2. QUANTUM MEASUREMENTS BASED ON SPINOR REPRESENTATIONS OF THE LORENTZ GROUP AND THEIR GENERALIZATIONS

The spinor representation of the Lorentz group is of fundamental importance for theoretical and mathematical physics. Our approach is based on the fact that a complex matrix $L$ of dimensions $2 \times 2$ with a determinant equal to one ( $\det(L) = 1$ ) generates a transformation that preserves the relativistic interval (Lorentz transformation) [1,2]. In paper [3], we considered the application of this formalism to the description of transformations and measurements of two-level quantum systems (qubits) and systems of higher dimensions. The purpose of this research is application of the results obtained in work [3] to the problem of precision control of the evolution of quantum states of multilevel systems (qudits). Next, we will show the essence of the spinor representation of the Lorentz group using the example of the mixed state of a qubit. Note that any initial mixed state of a qubit can be considered as a mixture of two different components.

---
[*] bogdanov_yurii@inbox.ru

Then, the mixed state, which we supplemented to a pure state, can be specified in the form of a complex matrix $\varphi_{in}$ of dimensions $2 \times 2$ (each column characterizes its component of the mixture). The original density matrix in this case is:

$$\rho_{in} = \varphi_{in} \cdot \varphi_{in}^+ \tag{1}$$

The procedure for adding a mixture to a pure state is ambiguous. In the general case, you can consider various complex matrices $\varphi_{in}$ of dimension $2 \times n$, $n \geq 2$ (a mixture of $n$ components). The matrix $\varphi_{in}$ can be multiplied on the right by an arbitrary unitary matrix $U$ of dimension $n \times n$: $\varphi_{in} \to \varphi_{in} U$. After transformation, the density matrix $\varphi_{in} \cdot \varphi_{in}^+$ obviously remains invariant. Note that the amplitude matrix $\varphi_{in}$ can be extended into a column of length amplitudes $2n$ (for this, the second column must be placed over the first, etc.).

However, for our purposes, it is more convenient to represent the pure state as a matrix $\varphi_{in}$ of dimensions $2 \times n$. The procedure for complementing a mixed state to a pure one plays a fundamental role in the problems of quantum measurements and tomography [4-5].

The spinor transformation is specified through a complex matrix $L$ of dimensions $2 \times 2$ with determinant equal to one:

$$\varphi_{out} = L\varphi_{in}, \quad \det(L) = 1 \tag{2}$$

The spinor transformation generates a density matrix at the output $\rho_{out} = \varphi_{out} \cdot \varphi_{out}^+$:

$$\rho_{out} = L \cdot \rho_{in} \cdot L^+ \tag{3}$$

This transformation preserves the determinant of the density matrix. This, in turn, ensures the invariance of the relativistic squared interval. Indeed, any qubit density matrix $\rho$ can be represented on the basis of four matrices ($I$, $\sigma_1$, $\sigma_2$, $\sigma_3$) [6]:

$$\rho = \frac{1}{2}\left(P_0 \cdot I + \vec{P}\vec{\sigma}\right) \tag{4}$$

The matrix $I$ is the identity matrix of dimension $2 \times 2$ and $\sigma_1, \sigma_2, \sigma_3$ are Pauli matrices. We will interpret this relationship in terms of quantum optics. In this case, the radiation intensity is $P_0 = Tr(\rho)$; and $P_1 = Tr(\rho\sigma_1)$, $P_2 = Tr(\rho\sigma_2)$, $P_3 = Tr(\rho\sigma_3)$ are the components of the Stokes vector. We emphasize that in the present consideration, the trace of the density matrix is not equal to unity, as is usually assumed, but to the radiation intensity $P_0$.

Using the standard representation for Pauli matrices, we obtain:

$$\rho = \frac{1}{2}\begin{pmatrix} P_0 + P_3 & P_1 - iP_2 \\ P_1 + iP_2 & P_0 - P_3 \end{pmatrix} \tag{5}$$

The invariance of the determinant of the density matrix for the spin transformation leads to the invariance of the value $s^2 = P_0^2 - \vec{P}^2$. The resulting value is the square of the relativistic interval. We consider the intensity $P_0$ as the zero component and the Stokes vector components $(P_0, \vec{P}) = (P_0, P_1, P_2, P_3)$ as the three spatial components of the four-Minkowski-Stokes's vector. We set the speed of light equal to one ($c = 1$). The Minkowski-Stokes four-vector can be interpreted as the four-momentum of some effective particle. It is easy to see that $\det(\rho) = \frac{1}{4}\left(P_0^2 - \vec{P}^2\right) = \frac{1}{4}m^2$, where $m$ is the mass of the considered effective particle. Note that the considered spinor representation of the Lorentz group is widely used in optical problems [7-11].

The speed of the effective particle $(P_0, \vec{P})$ corresponding to the four-momentum of Minkowski-Stokes with energy $P_0$ and momentum $\vec{P}$ is $\vec{v} = \vec{P}/P_0$. Let us consider the transformation of the transition to the frame of reference in which the momentum of the effective particle for the qubit vanishes (spatial components of the four-vector Minkowski-Stokes).

Such a frame of reference can be called the effective particle's own frame of reference (as well as the center of mass system). In this case, the qubit is in the center of the Bloch sphere. The transformation under consideration is described by the following pseudo-rotation:

$$L_{\vec{n}}(\theta) = \exp\left(-\frac{\theta}{2}\vec{\sigma}\vec{n}\right) = \cosh\left(\frac{\theta}{2}\right)I - \sinh\left(\frac{\theta}{2}\right)\vec{\sigma}\vec{n} \qquad (6)$$

Here $v = \tanh(\theta)$ is the modulus of the speed of the effective particle and $\vec{n}$ is the direction of the speed. The presented equation specifies the so-called Lorentz boost transformation, which describes the relative motion of reference frames at a constant speed without rotation of the coordinate axes. Note that a transition to an intrinsic frame of reference is possible only for a mixed state of a qubit. In the case of a pure state, such a transition is impossible, just as in relativistic physics it is impossible to go over to the photon's own frame of reference.

For example, we will use a polarization qubit. Let us consider the procedure for measuring the components of the four-vector Minkowski-Stokes. Let the light beam pass through a polarizer, the transmission axis of which is oriented vertically. $N_V$ is the number of photons that are transmitted by the polarizer for a time interval registered by detectors (the time interval we will conventionally take as a unit of time). In this case, the projection operator $|V\rangle\langle V|$ is measured. After rotating the polarizer by $90^0$, similar measurements of the projection operator will give the number of photons $N_H$. The total number of photons $N_V + N_H$ can serve as an estimate of the component $P_0 = Tr(\rho)$ of the four-vector Minkowski-Stokes, and the difference $N_V - N_H$ in the number of photons can serve as an estimate of the component $P_3 = Tr(\rho\sigma_3)$. We act by a unitary matrix $U$ on a photon before the polarizer. The transformed density matrix and the third component of the Stokes vector are: $\rho' = U\rho U^+$, $P_3' = Tr(U\rho U^+\sigma_3) = Tr(\rho U^+\sigma_3 U) = Tr(\rho\sigma_3')$, where $\sigma_3' = U^+\sigma_3 U$. If the polarizer is oriented vertically, then such a transformation leads to projection onto the state $U^+|V\rangle$ (in this case, the projection operator is measured as $U^+|V\rangle\langle V|U$). Similarly, with the horizontal orientation of the polarizer, the projection operator is measured as $U^+|H\rangle\langle H|U$. Obviously, the dimension of the projection $P_3$ corresponds to a unitary matrix equal to the identity matrix ($U_3 = I$). It is easy to see that the unitary rotations $U_1 = \frac{1}{\sqrt{2}}\begin{pmatrix} 1 & 1 \\ 1 & -1 \end{pmatrix}$ and $U_2 = \frac{1}{\sqrt{2}}\begin{pmatrix} 1 & -i \\ 1 & i \end{pmatrix}$ provide the measurement of the projections $P_1 = Tr(\rho\sigma_1)$ and $P_2 = Tr(\rho\sigma_2)$ of the four-vector of Minkowski-Stokes, respectively (we assume that $|V\rangle = \begin{pmatrix} 1 \\ 0 \end{pmatrix}, |H\rangle = \begin{pmatrix} 0 \\ 1 \end{pmatrix}$). Note that the considered unitary transformations provide the corresponding transformations of Pauli matrices, since $\sigma_1 = U_1^+\sigma_3 U_1, \sigma_2 = U_2^+\sigma_3 U_2$. Such transformations can be easily realized by means of phase plates.

The presented procedure for quantum measurements of a polarization qubit can be generalized to the case when it undergoes the Lorentz boost transformation before unitary rotations. It is especially important to consider the boost transformation corresponding to the transition to the system of the center of inertia for a qubit that is initially in an arbitrary mixed state. It turns out that in the system of the center of inertia, which corresponds to the qubit in the center of the Bloch sphere, the accuracy of quantum measurements is the highest (for a given number of measured representatives [5]). The quantum tomography protocol will be described by the so-called instrumental matrix $X$. In the simplest case, each row $Xj$ of this matrix defines a bra-vector in the Hilbert space onto which the measured state is projected, $j = 1, ..., m$, where m is the number of protocol lines (total number measured projections). The probability amplitude $M_j$ corresponding to the state vector $c$ and the given protocol line is $M_j = X_j c$ :. The probability per unit time $\lambda_j$, which specifies the expected number of registered events, is determined by the square of the modulus of the amplitude: $\lambda_j = |M_j|^2 = \langle c|\Lambda_j|c\rangle$, where $\Lambda_j = X_j^+ X_j$ is the measurement operator. If the state is described by the density matrix ρ, then $\lambda_j = \text{Tr}(\Lambda_j \rho)$.

Each line of the instrumental matrix is included in the protocol with its own weight. Let $t_j$ be the weight of the $j$-th row (exposure time). We will normalize the total weighted probability to the total sample size ($n$ is the total sample size for all lines of the protocol):

$$\sum_{j=1}^{m} t_j \lambda_j = n \qquad (7)$$

For protocols that are reduced to the decomposition of unity (the so-called POVM protocols), the sum of all protocol matrices is proportional to the identity matrix:

$$\sum_{j=1}^{m} t_j \Lambda_j = nI \qquad (8)$$

Here I is the identity matrix. Note that condition (8), together with the condition for normalizing the state vector, directly lead to (7).

As we showed in [3], the transformation of the original instrumental matrix $X^{in}$ under the action of the Lorentz transformation is reduced to multiplying it on the right by a complex matrix $L$:

$$X^{out} = X^{in} L \qquad (9)$$

We assume that all rows of the original instrumental matrix $X^{in}$ are normalized to one: $X_j^{in}(X_j^{in})^+ = 1$, $j = 1, 2, ..., m$. As a result of the Lorentz transformation with a non-unitary matrix $L$, the protocol line weights change. Let $t_j = X_j^{out}(X_j^{out})^+$ be the corresponding weights. They can be interpreted as the exposure times of the corresponding projection measurements (the exposure times of the lines of the original protocol $X^{in}$ are assumed to be equal to one). The accuracy of the reconstruction of the quantum state (Fidelity) is determined by the following equation [12,13]:

$$F = \left(Tr\sqrt{\sqrt{\rho}\rho_0\sqrt{\rho}}\right)^2 \qquad (10),$$

where $\rho_0$ is the theoretical density matrix, and $\rho$ is the reconstructed density matrix.

It can be shown [5] that for any POVM protocol the average accuracy loss $\langle 1-F \rangle$ cannot be lower than the following value:

$$\langle 1-F \rangle_{min} = \frac{v^2}{4n(s-1)} \qquad (11)$$

Here $v = (2s-r)r - 1$ is the number of degrees of freedom of the quantum state, $s$ is the dimension of the Hilbert space, $r$ is the rank of the density matrix, $n$ is the total sample size.

The efficiency of the protocol with respect to any considered quantum state can be defined as the ratio of the minimum possible loss of accuracy to the actually achievable one:

$$eff = \frac{\langle 1-F \rangle_{min}}{\langle 1-F \rangle} \qquad (12)$$

A fundamentally important point of our approach to constructing super-efficient quantum measurements is based on the fact that the Lorentz protocol is not reduced to the expansion of unity (POVM); therefore, the restriction (11) on the minimum loss of accuracy for it, generally speaking, is not necessary. It turns out that for this kind of measurements, it is possible that super-efficiency can arise, when formally $eff > 1$.

The Lorentz transformation considered above for an individual qubit can be generalized to the case of quantum systems of an arbitrary higher dimension. In this case, the transformation of the instrumental matrix into its own coordinate system is also carried out using the equation (9): $X^{out} = X^{in} L$.

Now the Lorentz transformation matrix is constructed according to the following rule [3]:

$$L = \frac{1}{\sqrt{s}} \psi_{in}^{-1}, \quad L \to \frac{L}{(\det L)^{1/s}} \qquad (13)$$

Note that the second relation defines a transformation that ensures the condition $\det L = 1$. Here $s$ is the dimension of the Hilbert space; $\psi_{in}$ - the purified state, which is given by the dimension $s \times s$ matrix and provides the following representation for the density matrix: $\rho_{in} = \psi_{in} \psi_{in}^+$

The matrix $L$ introduced by us specifies the generalized Lorentz transformation, which acts on the states given in $s$-dimensional Hilbert space. We do not associate this transformation with any tensor transformation in the usual four-dimensional space-time.

It can be seen that in the new coordinate system the density matrix is proportional to the unit matrix: $\rho_{out} = (\det \rho_{in})^{1/s} I$, where $I$ is the unit matrix of dimension $s \times s$. Note that of all mixed states, the highest accuracy is achieved precisely for such states, and not only for qubits, as mentioned above, but also for states of arbitrary dimension [5].

The example shown in Figure 1 corresponds to tomography of the quantum state of an eight-level qudit ($s=8$). The original measurement protocol specifies a set of MUB-mutually unbiased bases [14]. Figure 1 illustrates a close correspondence between the results of numerical experiments with the Lorentz protocol and the accuracy theory developed in our works [4-6,13].

The example presented in Figure 1 corresponds to a randomly generated mixed state of full rank ($r=8$), which is close to a pure state (the main component of the density matrix has a weight $\lambda_0 = 0.9999$). 200 numerical experiments were performed, the sample size (the total number of registered events is $n=10^4$) in each of the experiments was equal to the number of degrees of freedom $v=(2s-r)r-1=s^2-1=63$ of the quantum state. The average loss of accuracy, in this case, was extremely small $\langle 1-F \rangle = 2,15997 \cdot 10^{-6}$. As a result, in accordance with equation (12), the efficiency was very high $eff = 6563$.

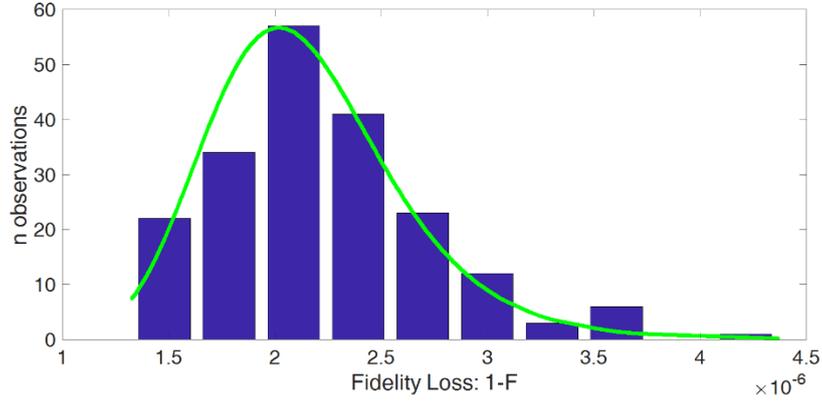

Figure 1. Distribution of fidelity loss for tomography of the quantum state of an eight-level qudit. Comparison of theory (curve) and numerical experiments (histogram).

## 3. ADAPTIVE TRACKING OF THE EVOLUTION OF THE QUANTUM STATE OF THE QUDIT

As noted above, the super-efficiency property of Lorentz protocols is based on the fact that these protocols do not reduce to unit expansions (do not form POVM). In fact, Lorentz's protocols describe unfinished measurements. This means that, after passing through a measuring system that meets the Lorentz protocol, a qudit (for example, ionic) will not necessarily be registered in one of the lines of the measurement protocol. Note that in the case of measuring ionic qudits, each elementary measurement "probes" the population at a separate level. In this case, in the case of Lorentz protocols, only a small part of the total number of ions is strongly perturbed (emit photons). The rest of the ions do not emit photons, but at the same time, they are still subjected to some final, albeit possibly weak, back action. It turns out that it is always possible to carry out the addition of the instrumental Lorentz matrix $X_L = X^{out}$ (see equation (9)) to the decomposition of unity (POVM). In this case, additional measurement operators can always be chosen so that they correspond to a weak disturbance of the initial state, while all the measurement operators of the original Lorentz protocol would set strong disturbances [3]. In the case of Lorentz protocols augmented to POVM, the overwhelming majority of the representatives of the quantum statistical ensemble are subject to only very weak perturbations. These representatives provide us with a resource for the procedure of tracking the evolution of a quantum state discussed below.

We assume that in the course of evolution, the quantum state of the system changes slowly, and that within one step of evolution, the quantum state does not change significantly (however, over a large number of steps, the quantum state changes radically).

The feedback algorithm is used to adapt the measurement protocol to the evolution of the quantum state in order to maintain a high accuracy of reconstruction of the changing state of the quantum system during the entire observation time. We used in our calculations the simplest algorithm for adapting the quantum measurement protocol, when the density matrix measured at the $j$-th step of evolution serves to construct the Lorentz protocol at the step $j+1$. At the same time, at the zero-evolution step, a standard MUB protocol based on mutually unbiased bases was used, which is not subject to the action of the Lorentz transformation (9).

The evolution of a quantum state at each individual step was determined by the infinitesimal unitary operator $U_0$:

$$\rho = U_0 \rho_0 U_0^{\dagger}, \text{где } U_0 = \exp(-i\varepsilon H) \tag{14}$$

Here $H$ is the evolution Hamiltonian, $\varepsilon$ is a small-time step.
We assume that the Hamiltonian $H$ can change slowly over time. In the course of modelling, we used the simplest periodic dependence of the Hamiltonian on time:

$$H = H_0 \left(1 + g \sin\left(\frac{2\pi j}{T}\right)\right) \tag{15}$$

Here $H_0$ is the unperturbed Hamiltonian (we chose a random Hermitian matrix of dimensions $s \times s$), $g$ is the parameter of the intensity of the periodic perturbation, $T$ is the period of oscillations and $j$ is the evolution step number.

As an example, we considered adaptive tomography of the evolving quantum state of an eight-level qudit ($s = 8$). As in the example in Figure 1, a measurement protocol was used based on a set of mutually unbiased bases, modified using the Lorentz transformation (9)). The simulation results are presented in Figures 2 and 3. The following parameters of the model were chosen: time step of evolution $\varepsilon = 3 \cdot 10^{-5}$; the parameter of the intensity of the periodic disturbance $g = 0.5$; the period $T$ was assumed to be equal to 1000 evolution steps ($T = 1000$).

The initial state was very close to the pure state (the main component of the density matrix had a weight $\lambda_0 = 0.999999$). Our goal was to control the quantum state with an accuracy of at least 4 nines (the degree of coincidence of the weakly perturbed and unperturbed states should be at least at the level $F = 0.9999$). For this, the Lorentz measurement protocol was tuned to measure the density matrix, the main component of which had a weight $\lambda_0 = 0.9999$ The sample size, which specifies the number of representatives registered in measurements, in each of the experiments was $n = 10^4$.

The protocol for measuring an eight-level qudit based on a set of mutually unbiased bases, modified using the Lorentz transformation, includes projecting into 72 different quantum states $|\varphi_j\rangle$, $j = 1, ..., 72$. The Lorentz protocols we are considering are aimed at measuring such states $|\psi\rangle$ for which the probabilities of registering events $|\langle\varphi_j|\psi\rangle|^2$ are small for all $j = 1, ..., 72$. In this case, the projector $|\varphi_j\rangle\langle\varphi_j|$ responds to the registration of the desired event and, at the same time, to a strong disturbance of the state. On the contrary, a projector $I - |\varphi_j\rangle\langle\varphi_j|$, where $I$ is the identity operator, corresponds to a qudit, which undergoes a very weak state disturbance and is not registered by the measurement system. Such qudits provide us with a resource for tracking the continuous tracking of the evolution of a quantum state. At the same time, we must continuously evaluate the back action of the measurements being carried out on the quantum system (qudit). To take into account such an impact at a separate step of evolution, instead of (14), we have the following transformation:

$$\rho = \left(I - |\varphi_j\rangle\langle\varphi_j|\right) U_0 \rho_0 U_0^\dagger \left(I - |\varphi_j\rangle\langle\varphi_j|\right) \tag{16}$$

We describe the degree of back action on weakly perturbed representatives through the degree of coincidence (Fidelity) of quantum states between the results of sequential action according to (14) for unperturbed representatives and the results of sequential action according to (16) for weakly perturbed representatives.

Figure 2 (top) shows that our adaptive method provides a reconstruction of the evolving quantum state of a qudit with an accuracy of more than 6 nines ($F > 0.999999$) The loss of accuracy, in this case, was an extremely small random variable with an average value $\langle 1 - F \rangle = 3,2637 \cdot 10^{-7}$ and standard deviation $\sigma_F = 4,2026 \cdot 10^{-8}$. Figure 2 (bottom) illustrates the high efficiency of the adaptive method for tracking the evolution of a quantum state. The efficiency (actually super-efficiency) of the protocol, in this case, was very high with the mean $\langle eff \rangle = 8403$ and standard deviation $\sigma_{eff} = 54,36$. We also see in this figure a weak periodic perturbation of the efficiency of quantum tomography with a period $T = 1000$, caused by the periodic modulation of the Hamiltonian.

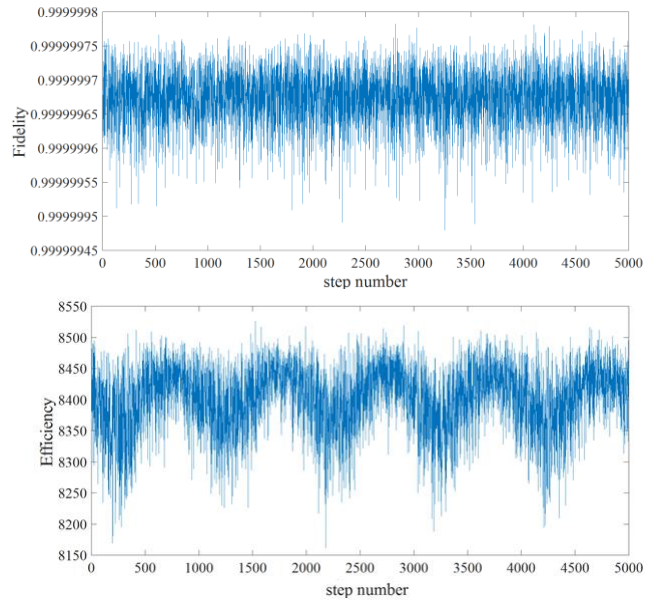

Figure 2. Evolution of the quality characteristic of tracking the quantum state of a qudit through weak measurements and feedback. Above - the fidelity of reconstruction of the quantum state, below - the efficiency (actually super-efficiency) of the protocol of quantum measurements

Figure 3 (top) shows the relative proportion of highly perturbed representatives for each of the 72 lines of the quantum measurement protocol. We see that the relative share of highly outraged representatives does not exceed $3 \cdot 10^{-5}$ for all lines of the protocol. On the contrary, Figure 3 (below) illustrates the degree of back action on slightly perturbed representatives (data from 5000 steps are presented in 30 grouping intervals). It turns out that under the influence of measurements, the degree of coincidence of the quantum states of unperturbed and weakly perturbed representatives already at the first step drops from 1 to the level of 0.99998 and then remains stable over time.

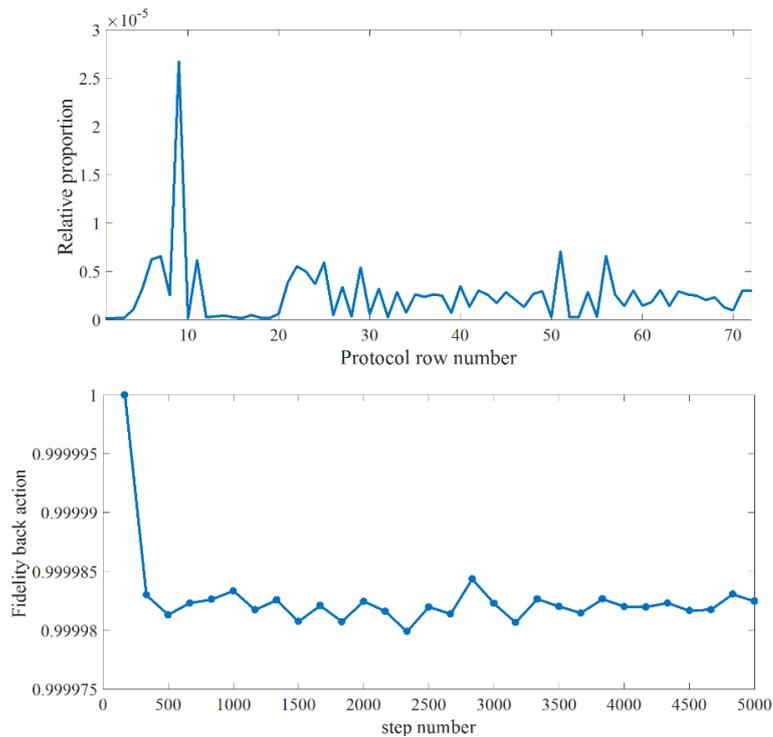

Figure 3. Estimation of the degree of the reverse effect of quantum measurements on the measured state. Above - the share of strongly perturbed representatives; below - the accuracy of the conservation of state (Fidelity) for weakly perturbed representatives

# CONCLUSIONS

The method of precision quantum tomography is proposed, which allows tracking with high accuracy the evolution of multilevel quantum systems (qudits) in Hilbert spaces of various dimensions. The developed quantum control algorithms are based on the use of the spinor representation of the Lorentz transformation group, as well as its generalizations to the case of multilevel quantum systems.

It is shown that feedback through weakly perturbing adaptive quantum measurements is capable of providing high-precision control of a quantum system, while introducing only weak perturbations into the initial quantum state. It turns out that, together with the control of a quantum system through its weak perturbation, the developed algorithms for controlling the evolution of the state of a quantum system can be super-efficient, providing a higher measurement accuracy than any standard POVM protocols.

The research results are important for the development of optimal adaptive methods for controlling quantum states and operations.

# ACKNOWLEDGMENTS

This work was supported by the Ministry of Science and Higher Education of the Russian Federation (program no. FFNN-2022-0016 for the Valiev Institute of Physics and Technology, Russian Academy of Sciences), and by the Foundation for the Advancement of Theoretical Physics and Mathematics BASIS (project no. 20-1-1-34-1)